\documentclass{article}
     \PassOptionsToPackage{numbers, compress}{natbib}

     \usepackage[final]{neurips_2020_ml4ps}

\usepackage[utf8]{inputenc} 
\usepackage[T1]{fontenc}    
\usepackage{hyperref}       
\usepackage{url}            
\usepackage{booktabs}       
\usepackage{amsfonts}       
\usepackage{nicefrac}       
\usepackage{microtype}      
\usepackage{amsmath}
\usepackage{graphicx}
\usepackage{subcaption}
\title{Better, Faster Fermionic Neural Networks}
\author{James Spencer$^\dagger$, David Pfau$^\dagger$, Aleksandar Botev$^\dagger$, W. M. C. Foulkes$^*$ \\
$^\dagger$DeepMind, $^*$Imperial College London \\
London, UK \\
\texttt{\{jamessspencer, pfau, botev\}@google.com}, \\ \texttt{wmc.foulkes@imperial.ac.uk}}

\begin{document}

\maketitle

\begin{abstract}
    The Fermionic Neural Network (FermiNet) \cite{pfau2020ab} is a recently-developed neural network architecture that can be used as a wavefunction Ansatz for many-electron systems, and has already demonstrated high accuracy on small systems. Here we present several improvements to the FermiNet that allow us to set new records for speed and accuracy on challenging systems. We find that increasing the size of the network is sufficient to reach chemical accuracy on atoms as large as argon. Through a combination of implementing FermiNet in JAX and simplifying several parts of the network, we are able to reduce the number of GPU hours needed to train the FermiNet on large systems by an order of magnitude. This enables us to run the FermiNet on the challenging transition of bicyclobutane to butadiene and compare against the PauliNet on the automerization of cyclobutadiene, and we achieve results near the state of the art for both.
\end{abstract}
\section{Introduction}

Accurate computational solutions to the Schr\"odinger equation are critical for quantum chemistry and condensed matter physics. The time-independent form of the equation tries to solve for a wavefunction $\psi(\mathbf{x}_1,\mathbf{x}_2,\ldots,\mathbf{x}_n)$, which is an eigenfunction of the Hamiltonian, $\hat{H}$ of the system:
\begin{eqnarray}
&\hat{H}\psi(\mathbf{x}_1,\ldots,\mathbf{x}_n) = E\psi(\mathbf{x}_1,\ldots,\mathbf{x}_n) 
\label{eqn:schrodinger}
\\
&\hat{H} = -\frac{1}{2}\sum_i \nabla^2_i + \sum_{i > j} \frac{1}{|\mathbf{r}_i-\mathbf{r}_j|} - \sum_{i I} \frac{Z_I}{|\mathbf{r}_i - \mathbf{R}_I|} 	+ \sum_{I > J} \frac{Z_I Z_J}{|\mathbf{R}_I-\mathbf{R}_J|}
\end{eqnarray}
Here $E$ is the real-valued energy of the wavefunction, $\mathbf{x}_i$ ($\mathbf{r}_i$) is the position and spin (position) of the $i$th electron, $\mathbf{R}_I$ is the position of the $I$th nucleus and $Z_I$ is the charge of the $I$th nucleus. Under the Born-Oppenheimer approximation, the nuclei are regarded as stationary and we can  solve for the lowest-energy solution of this equation by choosing a class of unnormalized wavefunctions parameterized by $\theta$, and minimizing the Rayleigh quotient:
\begin{align}
    \min_\theta \frac{\langle \psi_\theta \hat{H} \psi_\theta \rangle}{\langle \psi_\theta^2 \rangle} &= \min_\theta \frac{\int d\mathbf{x}_1\ldots d\mathbf{x}_n \psi_\theta(\mathbf{x}_1,\ldots,\mathbf{x}_n)\hat{H}\psi_\theta(\mathbf{x}_1,\ldots,\mathbf{x}_n)}{\int d\mathbf{x}_1\ldots d\mathbf{x}_n \psi_\theta^2(\mathbf{x}_1,\ldots,\mathbf{x}_n)} \nonumber \\
    &= \min_\theta \mathbb{E}_{\mathbf{x}_1,\ldots,\mathbf{x}_n \sim \psi^2_\theta}\left[\psi^{-1}(\mathbf{x}_1,\ldots,\mathbf{x}_n)\hat{H}\psi(\mathbf{x}_1,\ldots,\mathbf{x}_n)\right]
\end{align}
where in the last line we are performing the integral by Monte Carlo sampling from the distribution $p_\theta(\mathbf{x}) \propto \psi^2_\theta(\mathbf{x})$. This method of directly minimizing the energy of a system using a parametric approximation to the true wavefunction -- also known as a {\em wavefunction Ansatz} -- and Monte Carlo sampling is known as variational quantum Monte Carlo (VMC) \cite{foulkes2001quantum}.

 Many classes of wavefunction Ans\"atze have been proposed in the last fifty years. Deep neural networks have recently come to the fore as a promising flexible and expressive class of functions for the many-electron problem \cite{ruggeri2018nonlinear, luo2019backflow, han2019solving, choo2020fermionic, hermann2020deep}. The Fermionic Neural Network (FermiNet) \cite{pfau2020ab} has managed to reach higher absolute accuracy on many small atomic and molecular systems than other neural network Ans\"atze. This is likely due to a combination of architectural choices, and the use of Kronecker-Factored Approximate Curvature (KFAC) \cite{martens2015optimizing} as an optimizer, while other neural network Ans\"atze typically use first-order methods like ADAM. Some methods use the second-order stochastic reconfiguration \cite{sorella1998green} method, which is closely related to natural gradient descent, but are limited in their capacity to scale to larger networks.

Despite the promising FermiNet results, there are many limitations to the original work. Optimization requires large computational resources, and as system size grows, the accuracy diminishes slightly. In this paper, we investigate routes to improving the scaling, accuracy and optimization of the FermiNet, how to reach chemical accuracy on second row atoms, and how to accelerate the training of large systems by an order of magnitude.

\section{Improving FermiNet Accuracy}
\begin{table*}[t]
\centering
\begin{tabular}{c|cccc|c}\hline\hline
        \# hidden units & 256 & 512 & 256 & 512 & Exact \cite{chakravorty1993ground} \\
        \# dets & 16 & 16 & 32 & 32 & \\\hline
        P & -341.2561(1) & -341.2570(1) & -341.2565(1) & \textbf{-341.2578(1)} & -341.259 \\
        S & -398.1049(1) & -398.1066(1) & -398.1072(1) & \textbf{-398.1082(1)} & -398.110 \\
        Cl & -460.1452(1) & -460.1451(1) & -460.1463(1) & \textbf{-460.1477(1)} & -460.148 \\
        Ar & -527.5374(1) & -527.5384(1) & -527.5396(1) & \textbf{-527.5405(1)} & -527.540 \\\hline\hline
    \end{tabular}
    \caption{Accuracy comparison of FermiNets with different widths of one-electron streams and numbers of determinants for phosphorus through argon atoms. All energies are in Hartrees, $E_h$.}
    \label{tab:second_row_results}
\end{table*}

The original FermiNet achieved chemical accuracy, defined as 1 kcal/mol\footnote{1 kcal/mol is equal to 1.594 milli-Hartree (m$E_h$).} to the exact energy,  on many small systems, such as first-row atoms (lithium to neon), the accuracy declined with larger systems. Ablation studies on N$_2$, CO, and a hydrogen chain of 10 atoms showed that both the number of determinants and the width of the layers in the one-electron stream play an important role in the convergence of the FermiNet energy. We investigated the role of both in the performance on second-row atoms, specifically phosphorus through argon. While the settings used in the original paper were insufficient for reaching chemical accuracy relative to exact results \cite{chakravorty1993ground}, we found that increasing {\em both} the number of determinants and width of the one-electron stream was sufficient to reach chemical accuracy on all systems except sulphur. We also increased the number of MCMC steps between weight updates from 10 to 20 to reduce noise and improve equilibration, and ran training for 300,000-400,000 weight updates instead of 200,000 to guarantee convergence. Results are presented in Table~\ref{tab:second_row_results}.

\section{Scaling and Simplifying FermiNets}

\begin{table*}[t]
\centering
\begin{tabular}{cccccc}\hline\hline
        Framework & MCMC Steps & Det weights & Envelope & GPU Hours & Energy ($E_h$) \\\hline
        TensorFlow \cite{pfau2020ab} & 10 & Yes & Full Covariance & 11520 & -155.9263(6) \\
        JAX & 50 & No & Full Covariance & 1880 & \textbf{-155.9348(1)} \\
        JAX & 50 & No & Isotropic & \textbf{1104} & \textbf{-155.9348(1)} \\ \hline\hline
    \end{tabular}
    \caption{Speed and accuracy comparison for different training runs of the FermiNet on bicyclobutane. All runs were for 200,000 iterations, with the same number of determinants and hidden units as in \cite{pfau2020ab}. Number of GPU hours for TensorFlow assumes 30 days of training on 16 GPUs. Lower energies are strictly better as FermiNet provides an upper bound to the exact energy.}
    \label{tab:bicyclobutane_training}
\end{table*}

\begin{table*}[t]
\centering
\begin{tabular}{cccccc}\hline\hline
        Method & con\_TS & dis\_TS & g-but & gt\_TS & t-but \\\hline
        CCSD(T) \cite{kinal2007computational} & 40.4 & 21.8 & -25.1 & -22.3 & -28.0 \\
        CR-CC(2,3) \cite{kinal2007computational} & 41.1 & 66.1 & -24.9 & -22.1 & -27.9 \\
        CCSDt \cite{shen2012combining} & 40.1 & 59.0 & -27.2 & -25.3 & -31.1 \\
        CC(t;3) \cite{shen2012combining} & 40.2 & 60.1 & -25.3 & -22.6 & -28.3 \\
        DMC \cite{berner2010isomerization} & 40.4$\pm$0.5 & 58.6$\pm$0.5 & -25.2$\pm$0.5 & -22.2$\pm$0.5 & -27.9$\pm$0.5 \\ 
        FermiNet & 40.2$\pm$0.1 & 57.7$\pm$0.1 & -25.3$\pm$0.1 & -22.5$\pm$0.1 & -28.4$\pm$0.1 \\
        Experiment \cite{srinivasan1965thermal, wiberg1968heats} & 40.6$\pm$2.5 & - & - & - & -25.9$\pm$0.4 \\ \hline\hline
    \end{tabular}
    \caption{Comparison of the FermiNet against benchmark results on the transition from bicyclobutane to 1,3-butadiene. All energies are in kcal/mol and relative to the energy of bicylobutane calculated with the corresponding method. Energies are corrected for the zero-point vibrational energies \cite{kinal2007computational}.}
    \label{tab:bibut_to_butadiene}
\end{table*}

\begin{figure}
    \centering
    \begin{subfigure}[b]{0.45\textwidth}
    \includegraphics[width=\textwidth]{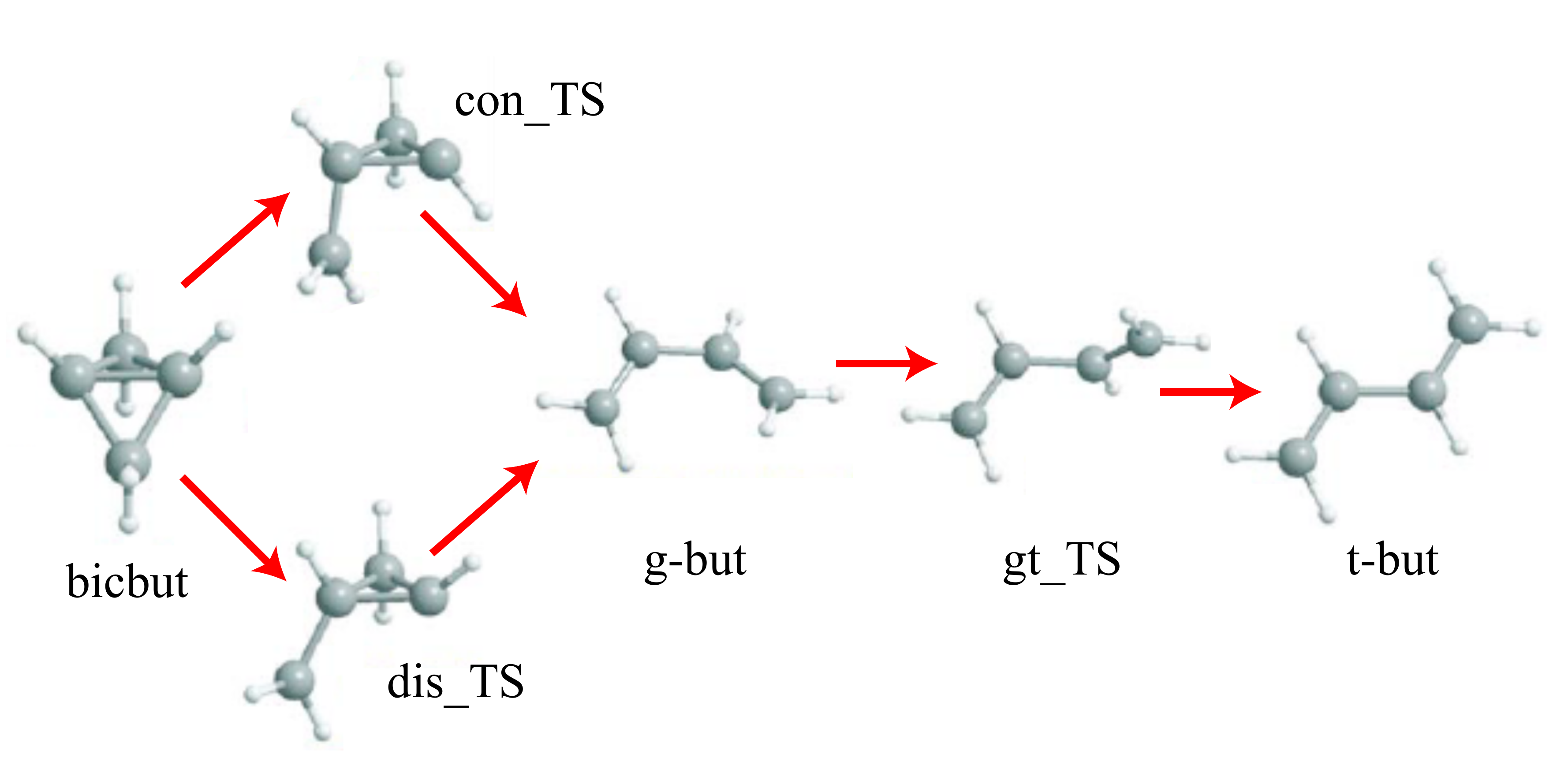}
    \vspace{0.5cm}
    \end{subfigure}~
    \begin{subfigure}[b]{0.55\textwidth}
    \includegraphics[width=\textwidth]{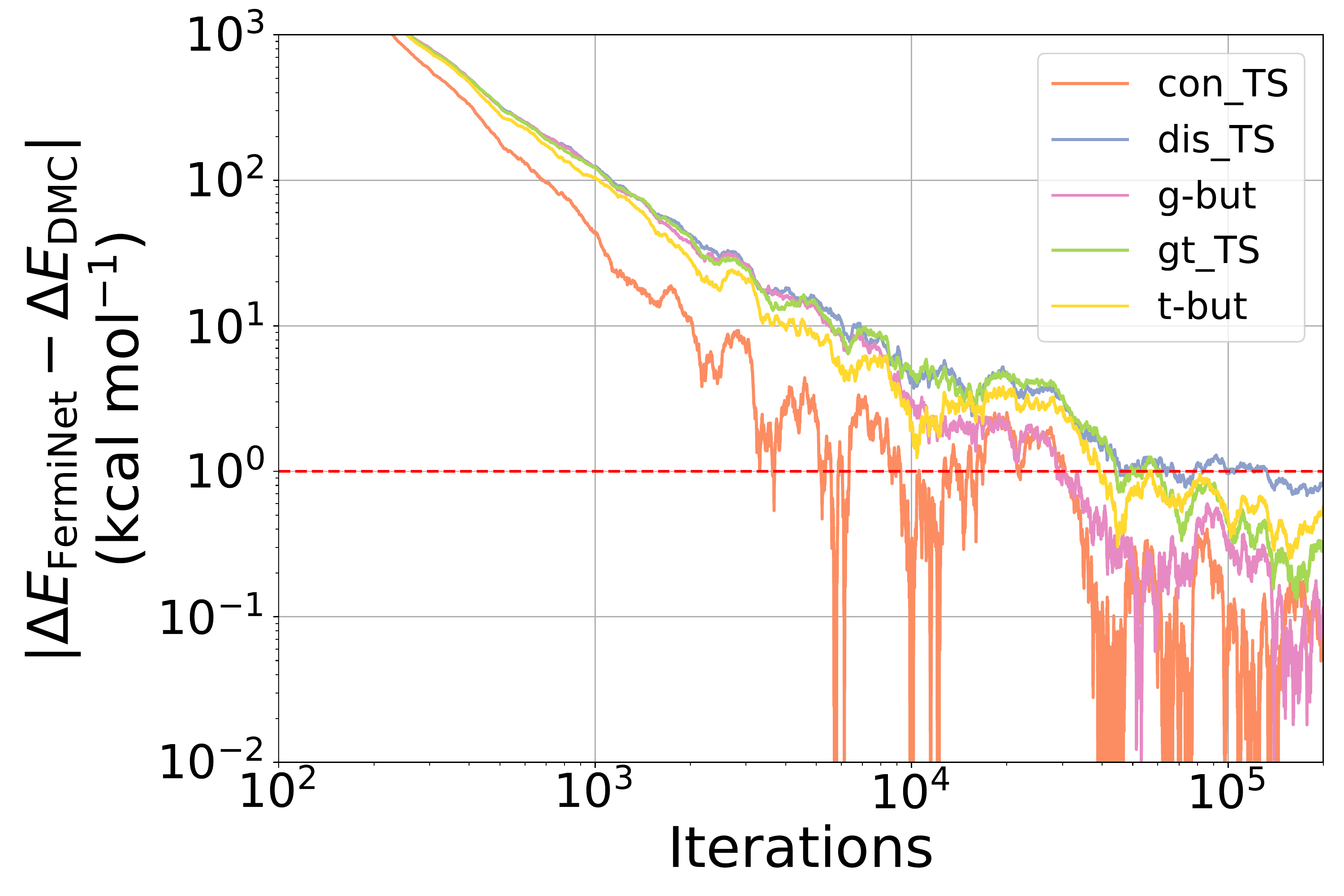}
    \end{subfigure}
    \caption{Bicyclobutane to 1,3-butadiene transition. Left: Illustration of the transition. There are two possible pathways -- conrotatory (con\_TS) and disrotatory (dis\_TS) -- for which CCSD(T) predicts dramatically different energies than other methods. 1,3-butadiene also has a \textit{gauche} (g-but) and \textit{trans} (t-but) isomer with an intermediate transition state (gt\_TS). Right: Learning curves for the FermiNet, giving the magnitude of the difference between the energy difference from bicyclobutane computed by the FermiNet ($\Delta E_{\mathrm{FermiNet}}$) and by DMC ($\Delta E_{\mathrm{DMC}}$). Energies are averages over the last 20\% of iterations for clarity. The energies have mostly converged after about 50,000 iterations.}
    \label{fig:bibut-pathway}
\end{figure}

The large computational overhead of the FermiNet is a limitation for practical adoption and scaling to larger systems. For instance, computing the energy of bicyclobutane, a 30-electron system, took roughly 1 month on 16 V100 GPUs \cite{pfau2020ab}. This is outside the reach of many research groups. We took several steps to improve this performance without sacrificing accuracy. First, we wrote a new implementation of the FermiNet in JAX \cite{bradbury2018jax}, which led to immediate improvements in performance. On large systems, GPU utilization went from \textasciitilde 60\% to \textasciitilde 90\%. The memory overhead also declined dramatically, possibly due to the availability of forward-mode gradients in JAX, which were used in computing the kinetic energy. This reduced the number of GPUs needed to run bicyclobutane with a batch size of 4096 from 16 to 4. On its own, this is enough to achieve a 6x improvement in efficiency.

To improve efficiency even further, we removed several unnecessary features from the FermiNet. At a high level, the FermiNet Ansatz can be written as:
\begin{equation}
    \sum_k \omega_k \mathrm{det}\left[\phi^{k\uparrow}_i\right] \mathrm{det}\left[\phi^{k\downarrow}_i\right]
\end{equation}
\begin{equation}
    \phi^{k\alpha}_i(\mathbf{r}^\alpha_j; \{\mathbf{r}^\alpha_{/j}\}; \{\mathbf{r}^{\bar{\alpha}}\}) =
    \left(\mathbf{w}^{k\alpha}_i \cdot \mathbf{h}^{L\alpha}_j + g^{k\alpha}_i\right)
	\sum_{m} \pi^{k\alpha}_{im}\mathrm{exp}\left(-|\mathbf{\Sigma}_{im}^{k \alpha}(\mathbf{r}^{\alpha}_j-\mathbf{R}_m)|\right)
	\label{eqn:envelope}
\end{equation}
Where $\alpha\in\{\uparrow, \downarrow\}$ index spins, $i$ and $j$ index electrons, $m$ indexes atoms and $k$ indexes determinants. $\mathbf{h}^{L\alpha}_j$ is the last layer of a permutation-equivariant neural network, and the last term in Eq.~\ref{eqn:envelope} involving weights $\pi^{k\alpha}_{im}$ and $\mathbf{\Sigma}_{im}^{k \alpha}$ is a multiplicative envelope that enforces the boundary condition that the wavefunction goes to zero at infinity. The weights $\omega_k$ on the determinants are redundant and can be absorbed into the linear weights $\mathbf{w}^{k\alpha}_i$, and so we remove them from the network. Each ``covariance" parameter $\mathbf{\Sigma}_{im}^{k \alpha}$ in the envelope is one $3\times 3$ matrix. Computing $|\mathbf{\Sigma}_{im}^{k \alpha}(\mathbf{r}^{\alpha}_j-\mathbf{R}_m)|$ adds quite a large computational overhead with unclear benefit. We find that if we replace the full covariance with a single parameter $\sigma_{im}^{k\alpha}$ and compute $|\sigma_{im}^{k \alpha}(\mathbf{r}^{\alpha}_j-\mathbf{R}_m)|$ instead, effectively making the covariance isotropic, the performance on bicyclobutane is unaffected, while the computational overhead is reduced by a full 40\%. In Table~\ref{tab:bicyclobutane_training}, it can be seen that by combining these simplications and the JAX implementation, training can be accelerated by a full order of magnitude, and the overall energy can actually be improved by a full 8 m$E_h$ relative to the original results. Note that this is {\em without} the wider networks and more determinants used in the previous section.

\paragraph{Bicyclobutane to 1,3-butadiene transition} The simplified FermiNet is fast enough that we can investigate multiple geometries of complex systems. We start with the transition of bicyclobutane to 1,3-butadiene. This has been investigated by many computational techniques \cite{kinal2007computational, shen2012combining, berner2010isomerization}, because the popular CCSD(T) method dramatically underestimates the energy of one possible transition, wrongly predicting that it is the preferred pathway. Both DMC and CC(t;3) methods agree with experimental results, and we compare against the DMC results in Fig.~\ref{fig:bibut-pathway} and the full suite of results in Table~\ref{tab:bibut_to_butadiene}. In both cases, we add the zero-point vibrational energy computed by CASSCF from \cite{kinal2007computational}. We find that with no fine tuning, the same FermiNet is able to match the DMC energy differences between bicyclobutane and all other states, both equilibrium and transition, to within chemical accuracy. In Table~\ref{tab:bibut_to_butadiene}, it can be seen that the FermiNet matches CC(t;3) even more closely, except on the disrotatory pathway, on which the FermiNet predicts slightly lower energies than any other accurate method. This is an impressive feat for a method applied essentially out-of-the-box, with no system-specific tuning.

\begin{figure}
    \centering
    \includegraphics[width=\textwidth]{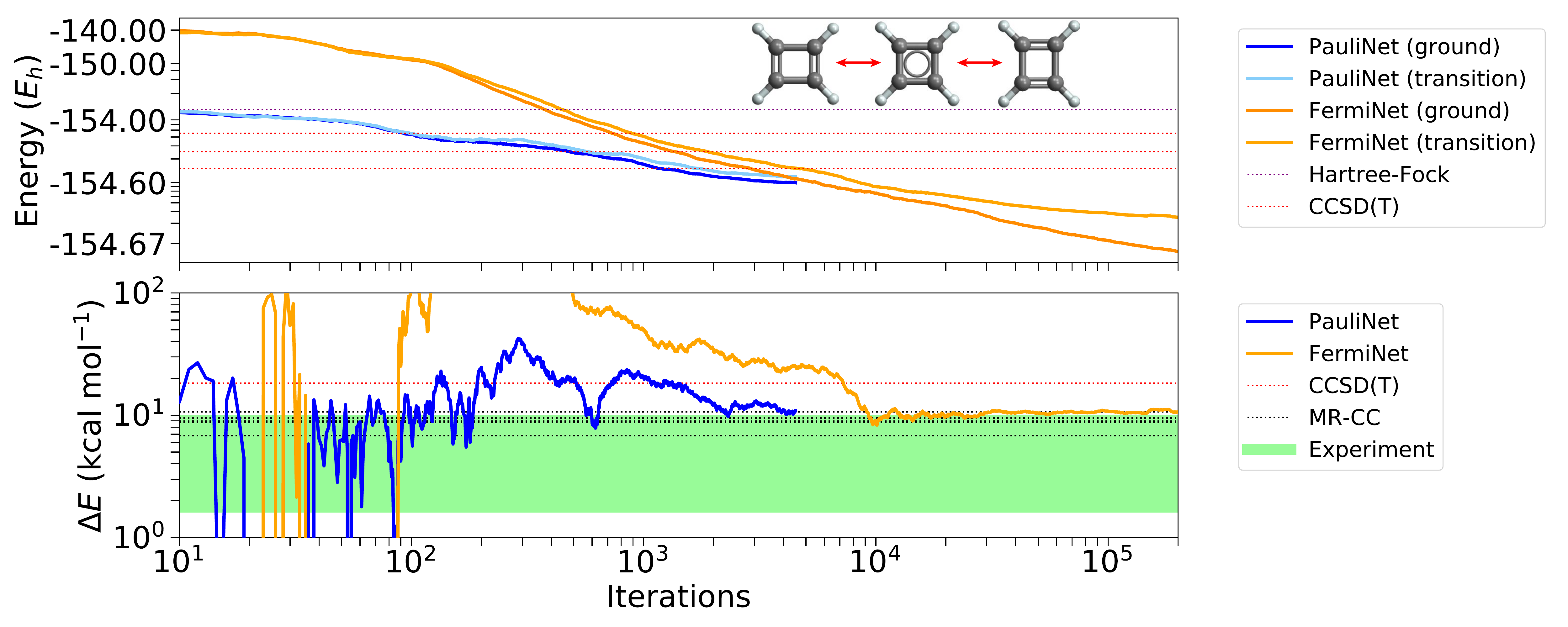}
    \caption{Automerization of cyclobutadiene and comparison with the PauliNet. The geometries of the ground state (left and right) and transition state (center) are shown in the top figure. Top: total energy of the FermiNet and PauliNet. Energy is plotted on a log scale zeroed at -154.68 E$_h$. While the FermiNet is initialized well above the PauliNet, the converged energy is \textasciitilde 70 m$E_h$ lower -- \textasciitilde 44 kcal/mol. Bottom: energy difference between the ground and transition state for the PauliNet and FermiNet. Both the PauliNet and FermiNet are at the very upper end of the experimentally-measured values. The FermiNet agrees well with the highest multireference coupled cluster (MR-CC) result. All results other than the FermiNet are from \cite{hermann2020deep}.}
    \label{fig:cyclobutadiene}
\end{figure}
\paragraph{Cyclobutadiene automerization and comparison to the PauliNet} The automerization of cyclobutadiene is another system on which CCSD(T) struggles, due to its multireference nature \cite{lyakh2012multireference}. The PauliNet, another neural network Ansatz which is faster but less accurate than the FermiNet, has been shown to match the performance of multireference coupled cluster on this system \cite{hermann2020deep}. We found that the FermiNet achieves energies \textasciitilde 70 m$E_h$ or 44 kcal/mol lower than the PauliNet on this system, but only after more iterations than the PauliNet was run for, so it is possible that the PauliNet did not fully converge. The relative energies between the ground and transition state were comparable between the PauliNet and FermiNet -- 9.9$\pm$0.6 kcal/mol for PauliNet and 10.3$\pm$0.1 kcal/mol for FermiNet -- and both were at the high end of the experimentally-observed range. In terms of the relative efficiency of the PauliNet and FermiNet, an exact comparison is difficult. Assuming full GPU utilization for both models, one iteration of the PauliNet took 50s on a GTX 1080 Ti, which at 11.3 TFLOP/s would come to 565 TFLOP/iteration. One iteration of the FermiNet took 2.5s across 8 V100s, which at 125 TFLOP/s per device comes out to 2.5 PFLOP/iteration, or roughly 5x the computational cost of the PauliNet. Already we have made impressive strides in accelerating the FermiNet, and we hope that future innovation will bring this figure down in the future.

\section{Conclusions}

We have shown that through a combination of careful engineering and simplification of the FermiNet, we can greatly increase the speed of training, while at the same time we can increase the accuracy simply by making the network larger. We can extend the range of atoms for which we can reach chemical accuracy to the entire second row of the periodic table, beyond which exact methods become impractical. While the original FermiNet results were not within chemical accuracy of CCSD(T) extrapolated to the complete basis set limit for large systems like bicyclobutane, we have shown here that the {\em relative} energies between different steps in the transition from bicyclobutane to butadiene are within chemical accuracy of the best available methods. This includes the disrotatory transition state, for which CCSD(T) fails dramatically. In comparison against the PauliNet, we find that the FermiNet is able to achieve far better {\em absolute} energies on cyclobutadiene, and comparable {\em relative} energies.

The only major downside of the FermiNet relative to the PauliNet is that it is significantly slower to train. One possible reason for this, advanced by the PauliNet authors, is that this is because the PauliNet incorporates more ``physical prior knowledge" -- namely the exact cusp conditions are built into the PauliNet, while the FermiNet learns the cusp conditions. Another possible reason is simply that the PauliNet sacrifices accuracy for speed, given it has an order of magnitude fewer parameters and does not achieve the same absolute energies as the FermiNet. The exact reason for the discrepancy in training speed remains a topic for future study. We are hopeful that future developments in neural network wavefunction Ans{\"a}tze will lead to models that combine the speed and light weight of the PauliNet with the accuracy of the FermiNet.

\section*{Broader Impact}

This work could lead to the adoption of new computational techniques by the chemistry and material science communities. While the present work is still too small to be applied to cutting-edge chemistry and has only been used for already well-understood systems, in the future these methods could help accelerate chemical research by predicting chemical reactions before they can be observed in the lab. In the best case, this could lead to discoveries with positive social impact like new lifesaving drugs or more efficient batteries or catalysts for carbon capture. In the worst case, this could lead to harmful discoveries of the sort that have happened in applied chemistry in the past -- but this is a risk shared by all computational chemistry methods development, and a strong professional ethos discouraging such research makes it unlikely.

\begin{ack}
Thanks to Piotr Piecuch and Frank No\'e for sharing data and results, and James Martens for discussions around KFAC and optimization.
\end{ack}

\bibliographystyle{plain}
\bibliography{main}

\begin{thebibliography}{10}

\bibitem{berner2010isomerization}
Raphael Berner and Arne L\"uchow.
\newblock Isomerization of bicyclo[1.1.0]butane by means of the {D}iffusion
  {Q}uantum {M}onte {C}arlo method.
\newblock {\em The Journal of Physical Chemistry A}, 114(50):13222--13227,
  2010.

\bibitem{bradbury2018jax}
James Bradbury, Roy Frostig, Peter Hawkins, Matthew~James Johnson, Chris Leary,
  Dougal Maclaurin, and Skye Wanderman-Milne.
\newblock {JAX}: composable transformations of {Python}+{NumPy} programs.
\newblock {\em \texttt{http://github.com/google/jax}}, 2018.

\bibitem{chakravorty1993ground}
Subhas~J Chakravorty, Steven~R Gwaltney, Ernest~R Davidson, Farid~A Parpia, and
  Charlotte~Froese p~Fischer.
\newblock Ground-state correlation energies for atomic ions with 3 to 18
  electrons.
\newblock {\em Physical Review A}, 47(5):3649, 1993.

\bibitem{choo2020fermionic}
Kenny Choo, Antonio Mezzacapo, and Giuseppe Carleo.
\newblock Fermionic neural-network states for ab-initio electronic structure.
\newblock {\em Nature Communications}, 11(1):1--7, 2020.

\bibitem{foulkes2001quantum}
WMC Foulkes, Lubos Mitas, RJ~Needs, and G~Rajagopal.
\newblock Quantum monte carlo simulations of solids.
\newblock {\em Reviews of Modern Physics}, 73(1):33, 2001.

\bibitem{han2019solving}
Jiequn Han, Linfeng Zhang, and E~Weinan.
\newblock Solving many-electron {S}chr{\"o}dinger equation using deep neural
  networks.
\newblock {\em Journal of Computational Physics}, 399:108929, 2019.

\bibitem{hermann2020deep}
Jan Hermann, Zeno Sch{\"a}tzle, and Frank No{\'e}.
\newblock Deep-neural-network solution of the electronic {S}chr\"odinger
  equation.
\newblock {\em Nature Chemistry}, 2020.

\bibitem{kinal2007computational}
Armaǧan Kinal and Piotr Piecuch.
\newblock Computational investigation of the conrotatory and disrotatory
  isomerization channels of bicyclo[1.1.0]butane to buta-1,3-diene: a
  completely renormalized coupled-cluster study.
\newblock {\em The Journal of Physical Chemistry A}, 111(4):734--742, 2007.

\bibitem{luo2019backflow}
Di~Luo and Bryan~K Clark.
\newblock Backflow transformations via neural networks for quantum many-body
  wave functions.
\newblock {\em Physical Review Letters}, 122(22):226401, 2019.

\bibitem{lyakh2012multireference}
Dmitry~I Lyakh, Monika Musia{\l}, Victor~F Lotrich, and Rodney~J Bartlett.
\newblock Multireference nature of chemistry: The coupled-cluster view.
\newblock {\em Chemical Reviews}, 112(1):182--243, 2012.

\bibitem{martens2015optimizing}
James Martens and Roger Grosse.
\newblock Optimizing neural networks with kronecker-factored approximate
  curvature.
\newblock In {\em International Conference on Machine Learning (ICML)}, pages
  2408--2417, 2015.

\bibitem{pfau2020ab}
David Pfau, James~S Spencer, Alexander~GDG Matthews, and W~Matthew~C Foulkes.
\newblock Ab initio solution of the many-electron {S}chr{\"o}dinger equation
  with deep neural networks.
\newblock {\em Physical Review Research}, 2(3):033429, 2020.

\bibitem{ruggeri2018nonlinear}
Michele Ruggeri, Saverio Moroni, and Markus Holzmann.
\newblock Nonlinear network description for many-body quantum systems in
  continuous space.
\newblock {\em Physical Review Letters}, 120(20):205302, 2018.

\bibitem{shen2012combining}
Jun Shen and Piotr Piecuch.
\newblock Combining active-space coupled-cluster methods with moment energy
  corrections via the {CC(P;Q)} methodology, with benchmark calculations for
  biradical transition states.
\newblock {\em The Journal of Chemical Physics}, 136(14):144104, 2012.

\bibitem{sorella1998green}
Sandro Sorella.
\newblock Green function monte carlo with stochastic reconfiguration.
\newblock {\em Physical review letters}, 80(20):4558, 1998.

\bibitem{srinivasan1965thermal}
R~Srinivasan, AA~Levi, and I~Haller.
\newblock The thermal decomposition of bicyclo[1.1.0]butane.
\newblock {\em The Journal of Physical Chemistry}, 69(5):1775--1777, 1965.

\bibitem{wiberg1968heats}
Kenneth~B Wiberg and Richard~A Fenoglio.
\newblock Heats of formation of {C$_4$H$_6$} hydrocarbons.
\newblock {\em Journal of the American Chemical Society}, 90(13):3395--3397,
  1968.

\end{thebibliography}

\end{document}